\documentclass[conference]{IEEEtran}
\ifCLASSINFOpdf
\else
\fi
\usepackage{amsmath}
\usepackage{graphicx}
\usepackage{epstopdf}
\usepackage{amssymb}
\usepackage{color}
\newcommand{\optval}[1]{#1^\ast}

\newcommand{\ookpulseprob}{q}
\newcommand{\Eul}{\mathrm{e}}
\begin{document}
%
% paper title
% Titles are generally capitalized except for words such as a, an, and, as,
% at, but, by, for, in, nor, of, on, or, the, to and up, which are usually
% not capitalized unless they are the first or last word of the title.
% Linebreaks \\ can be used within to get better formatting as desired.
% Do not put math or special symbols in the title.
\title{Efficiency of Optimized Pulse Position Modulation with Noisy Direct
Detection}

% author names and affiliations
% use a multiple column layout for up to three different
% affiliations
\author{
\IEEEauthorblockN{Marcin Jarzyna and Konrad Banaszek}
\IEEEauthorblockA{Centre of New Technologies\\
University of Warsaw\\
PL-02-097 Warszawa, Poland\\
E-mail: \{m.jarzyna, k.banaszek\}@cent.uw.edu.pl}}

% conference papers do not typically use \thanks and this command
% is locked out in conference mode. If really needed, such as for
% the acknowledgment of grants, issue a \IEEEoverridecommandlockouts
% after \documentclass

% for over three affiliations, or if they all won't fit within the width
% of the page, use this alternative format:
%
%\author{\IEEEauthorblockN{Michael Shell\IEEEauthorrefmark{1},
%Homer Simpson\IEEEauthorrefmark{2},
%James Kirk\IEEEauthorrefmark{3},
%Montgomery Scott\IEEEauthorrefmark{3} and
%Eldon Tyrell\IEEEauthorrefmark{4}}
%\IEEEauthorblockA{\IEEEauthorrefmark{1}School of Electrical and Computer Engineering\\
%Georgia Institute of Technology,
%Atlanta, Georgia 30332--0250\\ Email: see http://www.michaelshell.org/contact.html}
%\IEEEauthorblockA{\IEEEauthorrefmark{2}Twentieth Century Fox, Springfield, USA\\
%Email: homer@thesimpsons.com}
%\IEEEauthorblockA{\IEEEauthorrefmark{3}Starfleet Academy, San Francisco, California 96678-2391\\
%Telephone: (800) 555--1212, Fax: (888) 555--1212}
%\IEEEauthorblockA{\IEEEauthorrefmark{4}Tyrell Inc., 123 Replicant Street, Los Angeles, California 90210--4321}}

% use for special paper notices
%\IEEEspecialpapernotice{(Invited Paper)}

% make the title area
\maketitle

% As a general rule, do not put math, special symbols or citations
% in the abstract
\begin{abstract}
We analyze theoretically the impact of background counts on the
efficiency of optical communication in the photon-starved regime
using the pulse position modulation (PPM) format with direct
detection. Degradation of the photon information efficiency is
studied in the case when the background count rate is at most comparable
with the rate of photodetection events generated by the incoming
optical signal. The PPM symbol length is optimized under the
constraint of a fixed average spectral power density using an analytical
approximation. The results are
compared with generalized on-off keying (OOK) optimized over the a priori
probability distribution for the input binary alphabet. The generalized OOK scheme can be
viewed as a relaxation of the PPM scheme by removing the
requirement that a light pulse must occur exactly once in each
fixed-length frame of time bins that constitute the PPM symbol.
It is shown that the asymptotic scaling of the photon information
efficiency does not change qualitatively in this scenario.
\end{abstract}

% no keywords

% For peer review papers, you can put extra information on the cover
% page as needed:
% \ifCLASSOPTIONpeerreview
% \begin{center} \bfseries EDICS Category: 3-BBND \end{center}
% \fi
%
% For peerreview papers, this IEEEtran command inserts a page break and
% creates the second title. It will be ignored for other modes.
\IEEEpeerreviewmaketitle

\section{Introduction}

%{\color{blue} Objective: present simple formulas that can serve as estimates. PPM and OOK schemes - figures.}

Pulse position modulation (PPM) is the format of choice for deep-space optical communication, when the average signal power spectral density corresponds to much less than  the energy of one photon at the carrier frequency \cite{Hemmati2011}. In the PPM format the information is encoded in the position of a light pulse in a frame of otherwise empty time bins. It can be recovered on the receiver side from the timing of a detection event on a photon counting detector. A very weak light pulse may produce no photocount, resulting in the erasure of information encoded in a given frame. Erasures can be dealt with efficiently using forward error correction \cite{Dolinar2006}. The effective transmission rate depends on the PPM order, i.e.\ the number of time bins in one frame,  which should be adjusted to the available signal power \cite{Waseda2011}. A theoretical upper bound on the attainable transmission rate is given by Shannon mutual information. Under the constraint of a fixed average signal power spectral density, conventionally expressed in terms of the average photon number per time bin, the optimum of Shannon information with respect to the PPM order can be studied using analytical techniques \cite{Wang2014}, also taking into account super- or sub-Poissonian photon number statistics of light pulses employed in the communication link \cite{Jarzyna2015}.

In practice, intrinsic thermal excitations in the detector and stray optical radiation reaching the receiver also generate detection events \cite{Moision2014}. The purpose of this contribution is to discuss the efficiency of the PPM scheme in the presence of background noise. Specifically, we will study photon information efficiency when the background count probability is at most comparable with the average received number of photons in one time bin. We will show that in this regime one can obtain closed approximate formulas for the quantities of interest. These approximations are compared with the results of numerical optimization. Our analysis of the PPM scheme is restricted to the case of a simple-decision receiver, which interprets  as erasures instances when counts have been registered in two or more time bins within one PPM frame. In order to gain an insight into possible enhancement achievable through more sophisticated encoding and decoding strategies, we consider also generalized on-off keying (OOK) which uses a binary alphabet consisting of an empty time bin and a light pulse used with an arbitrary a priori probability distribution. The relative frequency with which pulses occur in the PPM format, given  by the inverse of the PPM order, can be related directly to the probability of sending a pulse in the generalized OOK scheme, but in the latter case it is no longer required that exactly one pulse must be sent in each frame of a constant length. Furthermore, Shannon mutual information for the generalized OOK scheme provides an upper bound on more intricate decoding schemes beyond the simple-decision model.

This paper is organized as follows. In Sec.~\ref{Sec:PPM} we review optimization of PPM with direct detection and no background noise. Results for generalized OOK without and with background noise are presented in Sec.~\ref{Sec:GOOK}. Optimization of PPM in the presence of background counts is discussed in Sec.~\ref{Sec:NoisyPPM}. Sec.~\ref{Sec:Conslusions} concludes the paper. The main results of the paper are analytical expressions (\ref{eq:IOOK}) and (\ref{eq:IPPM_noise}) for the photon information efficiency respectively in the case of noisy generalized OOK  and  noisy PPM with a simple-decision receiver. In the presented expressions $n_a$ is the average number of received photons per time bin and $n_b$ is the average number of background counts in one bin. Auxiliary functions used in these formulas are defined in Eqs.~(\ref{Eq:PIEPPMWW}) and (\ref{Eq:g(x)def}).

\begin{figure}[t]
\begin{centering}
\includegraphics[width=0.95\columnwidth]{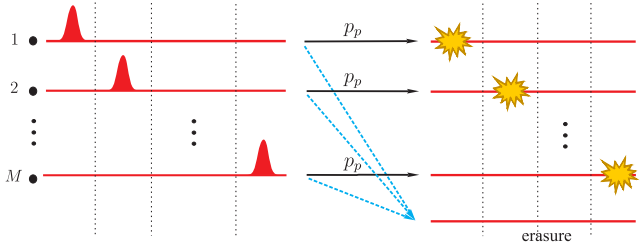}
\caption{PPM communication without background noise. The alphabet consists of $M$ symbols defined by the position of a light pulse in $M$ time bins. Direct detection identifies the symbol through the timing of a detector click with a probability $p_p$. In remaining instances the detector does not click at all resulting in an erasure outcome.}
\label{fig:scheme_PPM}
\end{centering}
\end{figure}

\section{PPM without background noise}
\label{Sec:PPM}

\begin{figure}[t]
\includegraphics[width=\columnwidth]{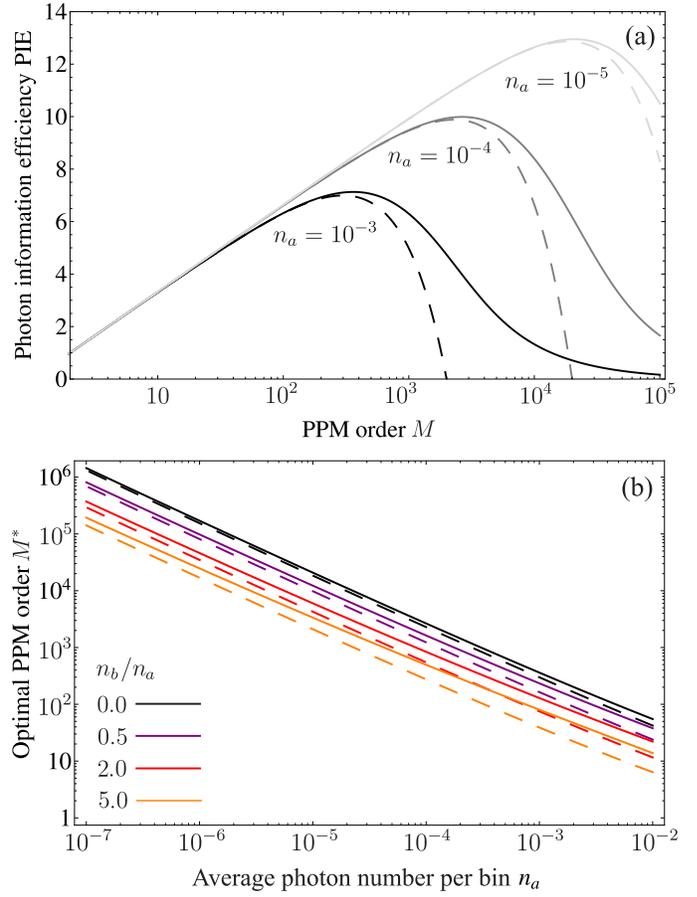}
\caption{(a) Photon information efficiency $\textrm{PIE}=I_\textrm{{PPM}}/n_a$ in the noiseless case as a function of the PPM order $M$ treated as a continuous variable for the average photon number $n_a=10^{-3}$ (black), $n_a=10^{-4}$ (dark grey), and $n_a=10^{-5}$ (light grey). Solid lines depict the exact expression, dashed lines are based on the second-order expansion (\ref{eq:p_s_approx}) of the photocount probability. (b) Numerically calculated optimal PPM order (black solid line) and the analytical approximation derived in Eq.~(\ref{eq:M_opt_apr}) (black dashed line) depicted as a function of the average photon number per time bin $n_a$. Color solid lines depict results of numerical optimization of the PPM order in the presence of background noise with the ratio $n_b/n_a$ of the average number of background counts $n_b$ to the average signal photon number per time bin $n_a$ specified in the legend. Color dashed lines represent respective results for the approximate expression given in Eq.~(\ref{eq:opt_M_apr_noise}).}
\label{fig:M_dependence}
\end{figure}

\begin{figure}
\includegraphics[width=\columnwidth]{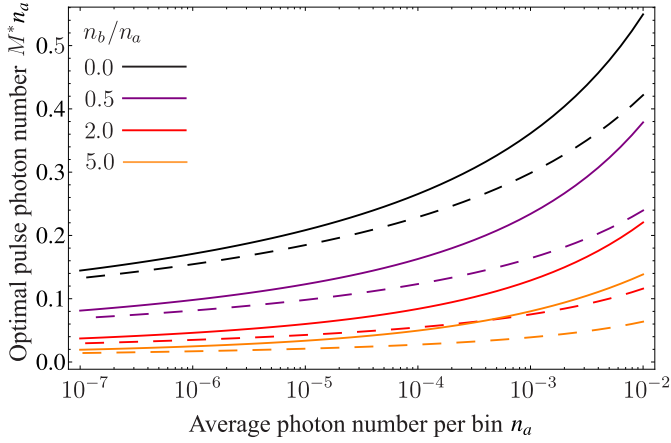}
\caption{Comparison of the optimal mean photon number in the received pulse $\optval{M}n_a$ calculated from the analytical approximation (dashed lines) with the results of numerical optimization (solid lines) in the noiseless case (black lines) and in the presence of background noise (color lines) with the ratio $n_b/n_a$ of the average number of background counts $n_b$ to the average signal photon number per time bin $n_a$ specified in the legend.}
\label{fig:Mna}
\end{figure}

Before addressing the impact of background counts, let us first review the efficiency of PPM communication in a background-free scenario. As depicted schematically in Fig.~\ref{fig:scheme_PPM}, the $M$-ary PPM format uses $M$ equiprobable symbols defined by the position of a single light pulse in a frame that consists of $M$ time bins. Direct detection of the incoming signal allows one to identify unambiguously the received symbol unless the detector does not click at all, which corresponds to an erasure outcome. For the average photon number per time bin in the received signal equal to $n_a$, the light pulse contains on average $M n_a$ photons which corresponds to the optical energy available in the entire frame under the average signal power constraint. Assuming Poissonian  statistics for the photon number distribution in the received light pulse, the probability that it will generate at least one photocount on the detector is $p_p = 1-\exp(-M n_a)$. We will refer to such an event as a detector click. To simplify subsequent formulas, the non-unit detection efficiency has been included in the overall transmission of the optical channel, lowering the effective value of $n_a$. Shannon mutual information expressed in bits per time bin for an $M$-ary erasure channel with non-erasure probability $p_p$ reads
\begin{equation}
\label{Eq:IPPMdef}
I_{\text{PPM}} = \frac{p_p}{M} \log_2 M.
\end{equation}
Anticipating that the optimal PPM order $M$ is high for $n_a \ll 1$, in the following we will treat $M$ as a continuous variable.
As seen in Fig.~\ref{fig:M_dependence}(a), for a fixed $n_a$ the mutual information $I_{\text{PPM}}$, or equivalently the photon information efficiency $\textrm{PIE}=I_{\text{PPM}}/n_a$, has a well defined maximum in $M$. Its approximate location can be found by assuming that the received pulse contains on average much less than one photon, $M n_a \ll 1$, and expanding the photocount probability $p_p$ up to the second order in $M n_{a}$,
\begin{equation}
\label{eq:p_s_approx}
p_p \approx M n_{a} - \frac{1}{2}  (M n_{a})^2 .
\end{equation}
Inserting this expansion into Eq.~(\ref{Eq:IPPMdef}) and equating to zero its first derivative with respect to $M$ yields a transcendental equation for the optimal $\optval{M}$, which can be solved in terms of the Lambert $W$ function \cite{Jarzyna2015}:
\begin{equation}\label{eq:M_opt_apr}
\optval{M} = \frac{2}{n_{a}} \left[ W \left( \frac{2 \Eul }{n_{a} } \right) \right] ^{-1}.
\end{equation}
The above approximation is compared with the results of numerical optimization of the exact expression for $I_{\text{PPM}}$ in Fig.~\ref{fig:M_dependence}(b). It is seen that the optimal PPM order is indeed high in the considered range of $n_a$ and that it grows monotonically with decreasing $n_a$.
Fig.~\ref{fig:Mna} depicts the mean photon number $\optval{M}n_a$ contained in the pulse for the optimal PPM order. Notably, over the range $10^{-7} \ll n_a \ll 10^{-2}$ covering five orders of magnitude, $\optval{M}n_a$ changes only by a factor of $4$.
The behavior in the limit $n_a \rightarrow 0$ can be obtained from the asymptotic form  \cite{Corless1996} of the Lambert function $W(x)$ for large arguments $x \gg 1$,
\begin{equation}\label{eq:lambert}
W(x) = \log x - \log\log x + \ldots,
\end{equation}
where the omitted terms are of the order in $x$ lower than a constant. Using the leading term of the above expansion
in Eq.~(\ref{eq:M_opt_apr}) yields
\begin{equation}
 \optval{M} n_{a} \approx 2\left[ \log \left( \frac{2 \Eul }{ n_{a} } \right) \right] ^{-1} ,
\label{Eq:Mna_approx}
\end{equation}
which implies that the mean photon number in the pulse decreases very slowly when $n_a$ tends to zero. In this regime Eq.~(\ref{Eq:Mna_approx}) implies the  hierarchy $n_a \ll  \optval{M} n_{a} \ll 1$, which justifies expansion (\ref{eq:p_s_approx}).

\begin{figure}
\includegraphics[width=\columnwidth]{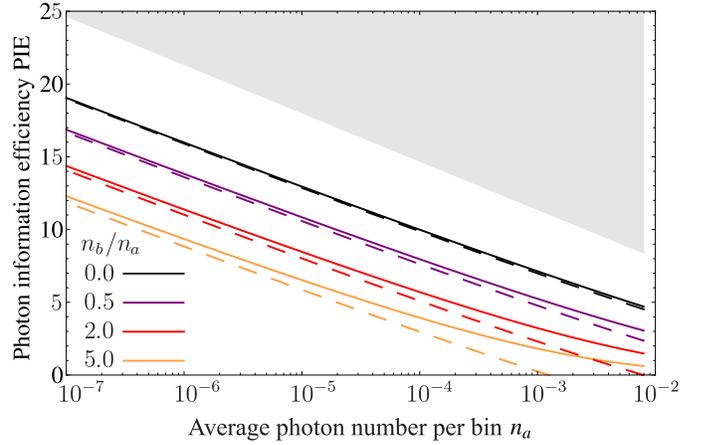}
\caption{Optimized photon information efficiency for a PPM communication link as a function of the average number of photons $n_a$ per time bin. Results of numerical optimization of the complete expression (solid lines) are compared with the approximate analytical formula (dashed lines). Black lines depict the case with no background noise, color lines are based on Eq.~(\ref{eq:PPM_noise}) for numerical optimization and on Eq.~(\ref{eq:IPPM_noise}) as the analytical approximation.
The shaded area represents the physically inaccessible region beyond the ultimate capacity of a lossy bosonic channel \cite{Giovannetti2004}.}
\label{fig:PPM}
\end{figure}

Inserting  the quadratic expansion (\ref{eq:p_s_approx}) and the approximate expression for the optimal PPM order (\ref{eq:M_opt_apr}) into Eq.~(\ref{Eq:IPPMdef}) yields a closed formula for the maximum Shannon information which can be cast as
$I_{\text{PPM}} = n_{a} \Pi (n_{a})$,
where $\Pi(n_a)$ is the photon information efficiency specifying how much information can be on average transmitted per one received photon for optimized PPM with direct detection and no background noise. This quantity can be written using the Lambert $W$ function in a compact form \cite{Kuszaj}
\begin{equation}
\Pi (n_{a}) =  \left\{ W \left( \frac{2\Eul}{n_{a}} \right) - 2
+ \left[ W \left( \frac{2\Eul}{n_{a}}\right) \right]^{-1} \right\} \log_2 \Eul.
\label{Eq:PIEPPMWW}
\end{equation}
In Fig.~\ref{fig:PPM} the above approximate formula is compared with results of numerical optimization of the complete expression given in Eq.~(\ref{Eq:IPPMdef}). The agreement is remarkably good.

The specific form of the photon information efficiency given in Eq.~(\ref{Eq:PIEPPMWW}) is convenient for analyzing the limit $n_a \rightarrow 0$. Using the asymptotics of the Lambert function (\ref{eq:lambert}) the expression given in Eq.~(\ref{Eq:PIEPPMWW}) can be expanded up to the  term constant in $n_a$ as
\begin{equation}
\Pi (n_{a}) = \log_2 \frac{1}{n_{a}} -  \log_2 \log \frac{2\Eul}{n_{a}} -\log_2 \frac{\Eul}{2} + \ldots.
\label{Eq:PIEPPMexp}
\end{equation}
This matches the results of the asymptotic study presented in \cite{Wang2014}.
It is instructive to compare this expression with the upper bound on the photon information efficiency implied by the ultimate capacity $C$ of a lossy bosonic channel under the constraint of a fixed mean photon number per channel use \cite{Giovannetti2004}, which in the limit $n_a \ll 1$ admits the expansion
\begin{equation}
\frac{1}{n_{a}}C  = \log_2 \frac{1}{n_{a}} + \log_2 \Eul + \ldots.
\label{Eq:PIECapExp}
\end{equation}
Both expressions (\ref{Eq:PIEPPMexp}) and (\ref{Eq:PIECapExp}) have the same form in the leading order of $n_a$, growing unbounded when $n_a \rightarrow 0$. A difference appears in the second to the leading order. The photon information efficiency for PPM includes a negative double logarithmic term $-\log_2\log({2\Eul}/{n_a})$ which produces a gap with respect to the ultimate capacity limit seen in Fig.~\ref{fig:PPM}.

%figure PPM
\section{Generalized on-off keying}
\label{Sec:GOOK}

\begin{figure}
\begin{centering}
\includegraphics[width=0.95\columnwidth]{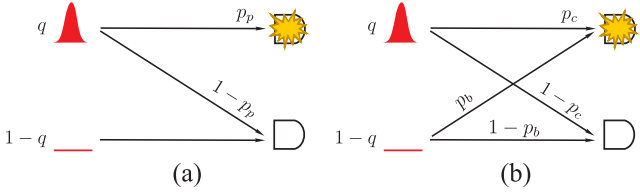}
\caption{Generalized on-off keying based on a binary alphabet composed of a light pulse and an empty bin used with respective probabilities $q$ and $1-q$. (a) In the absence of background noise an empty bin is identified unambiguously via direct detection while a light pulse is detected with a probability $p_p$. (b) When the background noise is present, an empty bin generates a click with a probability $p_b$ and a light pulse is detected with an overall probability $p_c$.}
\label{fig:scheme_GOOK}
\end{centering}
\end{figure}

\begin{figure}[t!]
\includegraphics[width=\columnwidth]{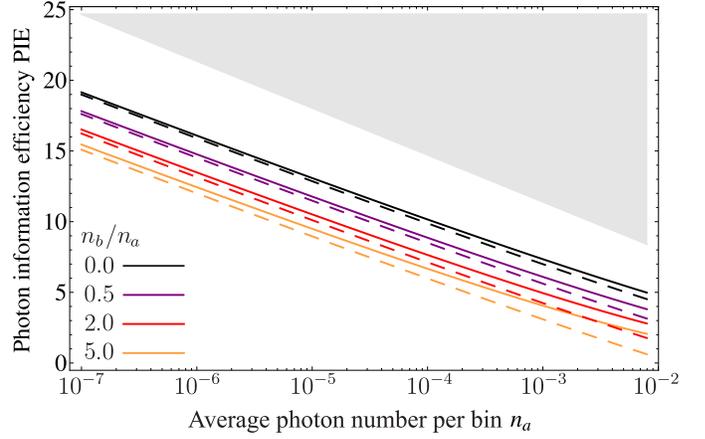}
\caption{Photon information efficiency as a function of the average number of photons $n_a$ per time bin for generalized OOK optimized over the a priori probability distribution for the input alphabet in the presence of background noise. Results of numerical optimization of the complete expression (solid lines) are compared with the approximate analytical formula (dashed lines). Black lines depict the case with no background noise, color lines are based on Eq.~(\ref{Eq:IGOOKnoisy}) for numerical optimization and on Eq.~(\ref{eq:IOOK}) as the analytical approximation. The shaded area represents the physically inaccessible region beyond the ultimate capacity of a lossy bosonic channel \cite{Giovannetti2004}.}
\label{fig:OOK}
\end{figure}

Let us now relax the requirement of the PPM format that exactly one pulse needs to be positioned in each frame of $M$ time bins. This is equivalent to generalized on-off keying, when light pulses are sent in arbitrarily chosen bins with the overall probability $\ookpulseprob$ and remaining bins are left empty. To satisfy the constraint on the average signal power spectral density, the mean photon number in a single pulse is $n_a/\ookpulseprob$. When noiseless direct detection is used at the output,  from the information theoretic viewpoint the scheme is described by the  $Z$ channel shown in Fig.~\ref{fig:scheme_GOOK}(a). An empty bin always produces an unambiguous outcome in the form of a no-count event, whereas a light pulse generates a detector click with a probability $p_p = 1-\exp(-n_a/\ookpulseprob)$, corresponding to registering at least one photocount. As before, Poissonian photon number statistics is assumed. Shannon mutual information for this scenario reads
\begin{equation}
I_{\text{OOK}} = H(\ookpulseprob p_p ) - \ookpulseprob H(p_p),
\label{Eq:IGOOKdef}
\end{equation}
where $H(x) =  - x \log_2 x - (1-x) \log_2 (1-x)$ is the binary entropy function. Let us now assume that both $\ookpulseprob$ and the mean photon number in the pulse $n_a/\ookpulseprob$ are much less than one. The expression for Shannon information can be recast to the form
\begin{multline}
I_{\text{OOK}} = \ookpulseprob p_p \log_2 \frac{1}{\ookpulseprob} + \ookpulseprob (1-p_p) \log_2 (1-p_p) \\
- (1- \ookpulseprob p_p) \log_2 (1- \ookpulseprob p_p).
\label{Eq:IGOOKexp}
\end{multline}
It is seen that the first term corresponds exactly to mutual information for the PPM scheme specified in Eq.~(\ref{Eq:IPPMdef}) with $\ookpulseprob = 1/M$, i.e.\ the relative frequency with which pulses occur in the PPM signal.
Maximization of the first term alone yields the same result as in the PPM scenario with the optimal $\optval{q} = 1/\optval{M}$. For this value, the combined second and third terms in Eq.~(\ref{Eq:IGOOKexp}) are of the order of $\optval{M} n_a^2$, hence their contribution to the photon number efficiency scales as $\optval{M} n_a$ which is lower order in $n_a$ than a constant according to Eq.~(\ref{Eq:Mna_approx}). Consequently it should be sufficient to retain only the first term in Eq.~(\ref{Eq:IGOOKexp}) and one can expect that the approximate formula derived in Eq.~(\ref{Eq:PIEPPMWW}) is applicable also to generalized OOK with direct detection and no background noise. This observation is confirmed in Fig.~\ref{fig:OOK}, which compares the photon information efficiency $I_{\text{OOK}}/n_a$ obtained from the numerical optimization of the complete expression given in Eq.~(\ref{Eq:IGOOKexp}) with Eq.~(\ref{Eq:PIEPPMWW}).

Let us now consider generalized OOK with direct detection in the presence of background noise. The corresponding channel model is shown schematically in Fig.~\ref{fig:scheme_GOOK}(b). We will assume that the background noise generates Poissonian count statistics with an average $n_b$. We will restrict our attention to the regime when the noise power does not exceed substantially the average signal power, i.e.\ the ratio $n_b/n_a$ is of the order of one. For an empty input bin a detector click generated by background noise occurs with a probability $p_b = 1 - \Eul^{-n_b}$, while the overall probability $p_c$ that an incoming light pulse results in a click is given by
\begin{equation}
p_c = p_p + p_b - p_p p_b = 1 - \exp(- n_b -n_a/\ookpulseprob ).
\label{Eq:pc=}
\end{equation}
In this setting Eq.~(\ref{Eq:IGOOKdef}) needs to be replaced by a more general expression for the Shannon mutual information which can be rearranged as
\begin{eqnarray}
I_{\text{OOK}} & = & H\bigl(\ookpulseprob p_c + (1-\ookpulseprob)p_b\bigr) - \ookpulseprob H(p_c) - (1-\ookpulseprob) H(p_b) \nonumber \\
&  = & \ookpulseprob p_c \log_2\frac{1}{\ookpulseprob}
+ (1-\ookpulseprob)p_b \log_2 \left(  \frac{p_b}{\ookpulseprob p_c} \right)
 \nonumber \\
& & - [\ookpulseprob p_c + (1-\ookpulseprob)p_b] \log_2 \left( 1 + \frac{(1-\ookpulseprob) p_b}{\ookpulseprob p_c} \right) \nonumber \\
& &  + (\ldots).
\label{Eq:IGOOKnoisy}
\end{eqnarray}
The first term in the second line has the form familiar from Eq.~(\ref{Eq:IGOOKexp}) with $p_p$ replaced by $p_c$ defined in Eq.~(\ref{Eq:pc=}). One can check that including background noise in $p_c$ does not change optimization noticeably. Hence the same approximate value for $\optval{\ookpulseprob}$ can be taken for the optimum as in the background-free case. Using this value we can insert approximations $\optval{\ookpulseprob} p_c \approx n_a$ and $p_b \approx (1-\optval{\ookpulseprob}) p_b \approx n_b$ in the second and the third term of Eq.~(\ref{Eq:IGOOKnoisy}). This allows us to write their sum as $n_a g(n_b/n_a)$, where
\begin{equation}
g(x) = (x+1) \log_2 (x+1) - x\log_2 x.
\label{Eq:g(x)def}
\end{equation}
Further, it can be verified that for the used $\optval{\ookpulseprob}$ the omitted terms denoted with $(\ldots)$ in Eq.~(\ref{Eq:IGOOKnoisy}) are of a lower order than $n_a$, therefore we will neglect them. Consequently one arrives at the following compact approximate formula for the generalized OOK photon information efficiency with background noise:
\begin{equation}\label{eq:IOOK}
{\text{PIE}}_{\text{OOK}} =  {\Pi} (n_{a}) -  g(n_b/n_{a}),
\end{equation}
where ${\Pi} (n_{a})$ is defined in Eq.~(\ref{Eq:PIEPPMWW}).
The derived approximation is compared with the results of numerical optimization in Fig.~\ref{fig:OOK}. It is seen that when the ratio
$n_b/n_a$ is fixed, the photon information efficiency exhibits the same unlimited growth with $n_a \rightarrow 0$ as in the noiseless case. In the approximate formula given in Eq.~(\ref{eq:IOOK}) the effect of background noise on the photon information efficiency is the subtraction of a constant term equal to $g(n_b/n_{a})$.

\section{PPM with noisy direct detection}
\label{Sec:NoisyPPM}

Let us now turn our attention to the case of PPM with noisy direct detection. The response of the detector in an individual time bin is given by the same set of conditional probabilities as for the generalized OOK scheme shown in Fig.~\ref{fig:scheme_GOOK}(b). An additional layer of complexity stems from the fact that in the presence of background noise clicks may occur in two or more distinct time bins within a single PPM frame. We will treat such instances as erasures and consider only outcomes when for a given PPM symbol the detector clicks at most in one time bin. The probability of the detector clicking exclusively in the bin corresponding to the position of the light pulse in the input symbol is
\begin{equation}
p_e = p_c (1-p_b)^{M-1} = \Eul^{-(M-1)n_b} - \Eul^{-M(n_a+ n_b)}.
\end{equation}
On the other hand, the probability that a click occurs in a specific time bin that differs from the position of the light pulse is
$p_d =  (1-p_c) p_b(1-p_b)^{M-2}$.
The full expression for the Shannon mutual information, expressed in bits per time bin, in such a scenario reads
\begin{multline}\label{eq:PPM_noise}
I_{\text{PPM}} = \frac{p_e}{M} \log_2 M  + \frac{M-1}{M} p_d \log_2 \frac{Mp_d}{p_e} \\
- \frac{p_e+(M-1)p_d}{M} \log_2 \left(1+\frac{(M-1)p_d}{p_e} \right).
\end{multline}
The right hand side has structure similar to that of Eq.~(\ref{Eq:IGOOKnoisy}). However, care needs to be taken when optimizing analytically the first term with respect to the PPM order $M$. This is because the quadratic expansion of the probability $p_e$ has the form
\begin{equation}
p_e \approx M n_{a} + \frac{\gamma}{2} (M n_{a})^2
\label{Eq:pe2ndorderexp}
\end{equation}
which compared to Eq.~(\ref{eq:p_s_approx}) differs in the second order by an additional factor $\gamma = 1 + 2 n_b/ n_{a}$. An analogous factor appears in the analysis of PPM efficiency in the absence of background noise, when the light pulse has non-Poissonian photon number statistics. In the latter case, $\gamma$ has the physical interpretation of the normalized second-order intensity correlation function of the light source \cite{Jarzyna2015}. Optimization of the first term in Eq.~(\ref{eq:PPM_noise})
using the expansion (\ref{Eq:pe2ndorderexp}) can be carried out similarly as in Sec.~\ref{Sec:PPM}, with the maximum value equal to $n_a \Pi (\gamma n_a)$ and attained at
\begin{equation}\label{eq:opt_M_apr_noise}
\optval{M} = \frac{2}{\gamma n_{a}} \left[ W \left( \frac{2 \Eul }{\gamma n_{a} } \right) \right] ^{-1}.
\end{equation}
This expression is compared with results of numerical optimization for several values of the ratio $n_b/n_a$ in Fig.~\ref{fig:M_dependence}(b).
The remaining two terms in Eq.~(\ref{eq:PPM_noise}) can be simplified using approximations $p_e \approx M n_a$ and $(M-1)p_d \approx M p_d \approx M n_b$. The final result for the photon information efficiency is
\begin{equation}\label{eq:IPPM_noise}
{\text{PIE}}_{\text{PPM}} =  {\Pi}(n_a + 2 n_b) -  g(n_b/n_{a}).
\end{equation}
The difference with the generalized OOK expression (\ref{eq:IOOK}) is in the first term. Because the function $\Pi$ is monotonically decreasing, adding a term $2n_b$ to its argument lowers its value. This effect is clearly seen in Fig.~\ref{fig:PPM} when compared to the generalized OOK case with the same ratio $n_b/n_a$ depicted in Fig.~\ref{fig:OOK}. The difference between ${\Pi}(n_a)$ and ${\Pi}(n_a + 2 n_b)$ can be estimated using the leading order of the expansion given in Eq.~(\ref{Eq:PIEPPMexp}). A simple calculation yields the value of the gap equal to $\log_2(1+2n_b/n_a)$. Consequently, for a fixed ratio $n_b/n_a$ the asymptotic scaling of the photon information efficiency is the same for both generalized OOK and PPM with a simple-decision receiver.

\section{Conclusions}
\label{Sec:Conslusions}

We have derived approximate analytical formulas for the photon information efficiency attainable using PPM and generalized OOK with noisy direct detection under the constraint of a given average power spectral density, expressed in terms of the average photon number per bin. The parameters of the input symbol ensemble have been optimized assuming that the ratio of the noise power to the average signal power is of the order of one. For the fixed ratio, with diminishing signal power the photon information efficiency in both scenarios follows in the leading order the ultimate capacity limit of a lossy bosonic channel. Remarkably, the performance of optimized PPM and generalized OOK is not significantly affected as long as the background noise power remains comparable to the average signal power.

The presented formulas can serve as a reference for benchmarking the performance of practical deep-space communication systems. It should be noted that with decreasing average power spectral density the optimal PPM order grows unbounded, which in the case of generalized OOK is reflected by the increasing imbalance between a priori probabilities for the input symbols. This results in escalating requirements for the peak-to-average power ratio, which may be difficult to meet with a satisfactory electrical-to-optical conversion efficiency.

% conference papers do not normally have an appendix

% use section* for acknowledgment
\section*{Acknowledgment}

The authors would like to thank to Saikat Guha and Piotr Kuszaj for insightful discussions. This work is part of the project ``Quantum Optical Communication Systems'' carried out within the TEAM
programme of the Foundation for Polish Science co-financed by the European Union under the European
Regional Development Fund.

% trigger a \newpage just before the given reference
% number - used to balance the columns on the last page
% adjust value as needed - may need to be readjusted if
% the document is modified later
%\IEEEtriggeratref{8}
% The "triggered" command can be changed if desired:
%\IEEEtriggercmd{\enlargethispage{-5in}}

% references section

% can use a bibliography generated by BibTeX as a .bbl file
% BibTeX documentation can be easily obtained at:
% http://mirror.ctan.org/biblio/bibtex/contrib/doc/
% The IEEEtran BibTeX style support page is at:
% http://www.michaelshell.org/tex/ieeetran/bibtex/
%\bibliographystyle{IEEEtran}
% argument is your BibTeX string definitions and bibliography database(s)
%\bibliography{IEEEabrv,../bib/paper}
%
% <OR> manually copy in the resultant .bbl file
% set second argument of \begin to the number of references
% (used to reserve space for the reference number labels box)
\IEEEtriggeratref{7}
\IEEEtriggercmd{\enlargethispage{-8.62in}}
\bibliographystyle{IEEEtran}
%\bibliography{Icsos_bib}

\begin{thebibliography}{9}
\providecommand{\url}[1]{#1}
\csname url@samestyle\endcsname
\providecommand{\newblock}{\relax}
\providecommand{\bibinfo}[2]{#2}
\providecommand{\BIBentrySTDinterwordspacing}{\spaceskip=0pt\relax}
\providecommand{\BIBentryALTinterwordstretchfactor}{4}
\providecommand{\BIBentryALTinterwordspacing}{\spaceskip=\fontdimen2\font plus
\BIBentryALTinterwordstretchfactor\fontdimen3\font minus
  \fontdimen4\font\relax}
\providecommand{\BIBforeignlanguage}[2]{{%
\expandafter\ifx\csname l@#1\endcsname\relax
\typeout{** WARNING: IEEEtran.bst: No hyphenation pattern has been}%
\typeout{** loaded for the language `#1'. Using the pattern for}%
\typeout{** the default language instead.}%
\else
\language=\csname l@#1\endcsname
\fi
#2}}
\providecommand{\BIBdecl}{\relax}
\BIBdecl

\bibitem{Hemmati2011}
H.~Hemmati, A.~Biswas, and I.~B. Djordjevic, ``Deep-space optical
  communications: Future perspectives and applications,'' \emph{Proc. IEEE},
  vol.~99, no.~11, pp. 2020--2039, nov 2011.

\bibitem{Dolinar2006}
S.~J. Dolinar, J.~Hamkins, B.~E. Moision, and V.~A. Vilnrotter,
  \emph{Deep-Space Optical communications}.\hskip 1em plus 0.5em minus
  0.4em\relax New York, NY: Wiley, 2006, ch.~4.

\bibitem{Waseda2011}
A.~Waseda, M.~Sasaki, M.~Takeoka, M.~Fujiwara, M.~Toyoshima, and A.~Assalini,
  ``Numerical evaluation of PPM for deep-space links,'' \emph{J. Opt. Commun.
  Netw.}, vol.~3, no.~6, pp. 514--521, jun 2011.

\bibitem{Wang2014}
L.~Wang and G.~W. Wornell, ``A refined analysis of the Poisson channel in the
  high-photon-efficiency regime,'' \emph{IEEE Trans. Inf. Theor.}, vol.~60,
  no.~7, pp. 4299--4311, jul 2014.

\bibitem{Jarzyna2015}
M.~Jarzyna, P.~Kuszaj, and K.~Banaszek, ``Incoherent on-off keying with
  classical and non-classical light,'' \emph{Opt. Express}, vol.~23, no.~3, pp.
  3170--3175, feb 2015.

\bibitem{Moision2014}
B.~Moision and W.~Farr, ``Range dependence of the optical communications
  channel,'' \emph{IPN Prog. Rep.}, pp. 42--199, nov 2014.

\bibitem{Corless1996}
R.~M. Corless, G.~H. Gonnet, D.~E.~G. Hare, D.~J. Jeffrey, and D.~E. Knuth,
  ``On the Lambert W function,'' \emph{Adv. Comput. Math.}, vol.~5, no.~4, pp.
  329--359, dec 1996.

\bibitem{Kuszaj}
P.~Kuszaj, private communication.

\bibitem{Giovannetti2004}
V.~Giovannetti, S.~Guha, S.~Lloyd, L.~Maccone, J.~H. Shapiro, and H.~P. Yuen,
  ``Classical capacity of the lossy bosonic channel: the exact solution,''
  \emph{Phys. Rev. Lett.}, vol.~92, no.~2, p. 027902, jan 2004.

\end{thebibliography}
%\end{document}

\end{document}